\documentstyle[prl,multicol,aps,tighten]{revtex}
\def\bea{\begin{eqnarray}}
\def\eea{\end{eqnarray}}
\def\be{\begin{equation}}
\def\ee{\end{equation}}
\begin{document}


\title{{\bf Scale Vs. Conformal Invariance in the AdS/CFT Correspondence}}

\author{Adel M. Awad{$^\sharp$} 
and Clifford V. Johnson$^\natural$}

\address{\hfil}
\address{$^\sharp$Department of Physics and Astronomy,
University of Kentucky,  Lexington, KY 40506, U.S.A.\\and\\
Department of Physics, 
Faculty of Science, Ain Shams University, Cairo 11566, Egypt}
\address{$\phantom{and}$}
\address{$^\natural$Centre 
for Particle Theory, Department of Mathematical Sciences, University of
Durham, Durham, DH1 3LE, U.K.}
\address{\hfil\\\tt adel@pa.uky.edu, c.v.johnson@durham.ac.uk}

\address{\hfil}

\date{June 2000}
\maketitle

\begin{abstract}
We present two examples of non--trivial field theories which are scale
invariant, but not conformally invariant. This is done by placing
certain field theories, which are conformally invariant in flat space,
onto curved backgrounds of a specific type.  We define this using the
AdS/CFT correspondence, which relates the physics of gravity in
asymptotically Anti--de Sitter (AdS) spacetimes to that of a conformal
field theory (CFT) in one dimension fewer. The AdS rotating (Kerr)
black holes in five and seven dimensions provide us with the examples,
since by the correspondence we are able to define and compute the
action and stress tensor of four and six dimensional field theories
residing on rotating Einstein universes, using the ``boundary
counterterm'' method. The rotation breaks conformal but not scale
invariance.  The AdS/CFT framework is therefore a natural arena for
generating such examples of non--trivial scale invariant theories
which are not conformally invariant.
\end{abstract}

\begin{multicols}{2}

There is an often quoted piece of folklore in the subject that states
that any non--trivial example of a field theory which is scale
invariant is automatically conformally invariant. Proofs of this
statement only exist in certain specific situations, and in fact it is
known to be not generally applicable. Some discussion can be
found in {\it e.g.,} refs.\cite{jackiwone,jackiwtwo,paul,joe,ivo}. 
 The focus in those cases is on
finding field theory examples in flat space. Another way to violate
conformal invariance is of course to place the field theory on a
spacetime with non--zero curvature. Then there are conformal
anomalies, and the non--vanishing trace of the stress--energy tensor,
${\widehat T}_{ab}$, can be written in terms of various local
curvature invariants of the background spacetime.  The general form of
the anomaly in dimension $n$ (which is even, since we only have
conformal anomalies in those cases) is given by (see {\it e.g.}
ref.\cite{deser}): \be {\widehat T}_a^a= {\rm c}_0 E_n + \sum_i {\rm
c}_i I_i + \nabla_a J^a\ . \ee Here, the ${\rm c}$'s are constants,
$E_n$ is the Euler density, $I_i$ are terms constructed from the Weyl
tensor and its derivatives, and the last term is a collection of total
derivative terms. The first type of term is called ``type~A'', the
next ``type~B'', and the last ``type~D''.  It is important to note
that the coefficients of all terms are regularisation scheme
independent {\sl except} the type~D anomaly. These latter terms can be
removed by a suitable addition of local counterterms to the
action. The type~A anomaly is only locally a total derivative, in
general.

In order to construct a non--trivial example of a scale invariant
theory which is not conformally invariant, we can simply place a
conformally invariant theory on a spacetime $\cal S$ for which the type A
anomaly does not vanish, while the types B and D anomalies do. In this
case, there will be an irremovable anomaly ${\widehat T}^a_a$, for
which $\int_{\cal S}\, {\widehat T}^a_a=0$, showing that scale invariance
is preserved.

In short, we must find a way to place a conformally invariant theory
on a spacetime for which the Euler density is non--vanishing, but
which is topologically trivial, so that the integral vanishes. Our
spacetime must also be conformally flat, thus not contributing to the
Type~B anomaly, which would break scale invariance too. In this
letter, we show how to do this, and in this way find a new class of
counterexamples to the folklore.

New tools have appeared on the market for defining and studying
conformally invariant field theories in interesting situations, often
even at strong coupling. One of these, the ``AdS/CFT correspondence'',
relates an $(n{+}1)$--dimensional theory of gravity on anti--de Sitter
(AdS) spacetime (times a compact manifold) to a conformal field theory
(CFT) in $n$ dimensions. This duality  arose as a result of
investigating\cite{Maldacena} a large number, $N$, of parallel D3--branes (reviewed in
refs.~\cite{notes}) in the context of the low energy, classical limit
of type~IIB superstring theory, ---the supergravity limit--- on five
dimensional anti--de Sitter spacetime times a five sphere
(AdS$_5{\times}S^5$). The dual CFT in this case is the four
dimensional ${\cal N}{=}4$ supersymmetric $SU(N)$ Yang--Mills theory
for large $N$. For other dimensions, the dual theories exist, but are
less well understood. For example eleven dimensional supergravity on
AdS$_7{\times}S^4$ is dual to a six dimensional ``(0,2)'' CFT, (the
notation denoting the number and chirality of the six dimensional
supercharges), also at large $N$. (See ref.\cite{review} for a review.) A precise
statement of the AdS/CFT correspondence\cite{Witten,Gubser} equates
the partition functions: \bea Z_{AdS}(\phi_{i})=Z_{CFT}(\phi_{0,i})\ .
\eea {}From the gravity--on--AdS point of view, $\phi_{i}$ is a bulk
field constrained to the values $\phi_{0,i}$ on the AdS boundary,
while from the CFT point of view, $\phi_{0,i}$ are sources for
pointlike operators, ${\cal O}_i$, in the theory. In the low energy
limit of the theory one can use the classical gravitational action to
calculate the partition function of the CFT ``on the boundary''. This
action has the form\cite{Gibbons}, \bea I_{\rm bulk}+I_{\rm surf}=
&-&{1 \over 16 \pi G}\int_{\cal M} d^{n+1}x \sqrt{-g}\left(R+{n(n-1)
\over l^2}\right)\nonumber\\ &-&{1 \over 8 \pi G } \int_{\partial
{\cal M}} d^{n}x \sqrt{-h} \,K.  \eea The first term is the
Einstein--Hilbert action with negative cosmological constant
($\Lambda{=}{-n(n{-}1)/2l^2}$). The second term is the
Gibbons--Hawking boundary term.  Here, $h_{ab}$ is the boundary metric
and $K$ is the trace of the extrinsic curvature $K^{ab}$ of the
boundary.

The theory on the boundary can be seen to obtain its conformal
invariance properties from two (related) sources: First, the metric on
the boundary of the theory is not uniquely defined, since the AdS
metric has a double pole there. It is instead only defined up to a
conformal class of metrics. The double pole divergence of the metric
is precisely what allows the theory to inherit the properties of a
conformal field theory, since the pole shows up as the correct
behaviour of the operator product expansion in the
CFT\cite{Witten}. Second, the $SO(n{+}1,2)$ isometry of the
AdS$_{n+1}$ spacetime descends to the local conformal group of the
field theory on the boundary\cite{Maldacena}.

The AdS/CFT correspondence is therefore a powerful way of studying
properties of the dual conformal field theory, by relating them to
properties of the gravity theory.  Here, we will compute the action
and stress tensor for certain gravity solutions and relate them to
properties of their dual field theories.  To deal with the divergences
which appear in the gravitational action (arising from integrating
over the infinite volume of spacetime), we shall use the ``counterterm
subtraction'' method\cite{Balasubramanian}, which regulates the action
by the addition of certain boundary counterterms which depend upon the
geometrical properties of the boundary of the spacetime. They are
chosen to diverge at the boundary in such a way as to cancel the bulk
divergences\cite{Balasubramanian,counter} (see also
refs.\cite{Henningson,lau,mann,kraus,solo}): \bea &&I_{\rm ct}={1
\over 8 \pi G} \int_{\partial {\cal M}}d^{n}x\sqrt{-h}\Biggl[
\frac{(n-1)}{ l}-{l{\cal R} \over 2(n-2)}+\nonumber\\
&&\qquad{l^3\over2(n-4)(n-2)^2}\left({\cal R}_{ab}{\cal
R}^{ab}-{n\over 4(n-1)}{\cal R}^2\right)\Bigg]\ .
\label{theterms}
\eea Here ${\cal R}$ and ${\cal R}_{ab}$ are the Ricci scalar and
tensor for the boundary metric $h$.  Using these counterterms one can
construct a divergence--free stress tensor from the total action
$I{=}I_{\rm bulk}{+}I_{\rm surf}{+}I_{\rm ct}$ by defining (see {\it
e.g.}  ref.\cite{Brown}): \bea T^{ab}&=& {2 \over \sqrt{-h}} {\delta I
\over \delta h_{ab}}\ .
\label{stressone}
\eea

For orientation, a metric on AdS$_{n+1}$, in global coordinates is:
\be ds^2=-\left(1+{r^2\over l^2}\right)dt^2
+{dr^2\over\left(1+{r^2/l^2}\right)}+r^2d\Omega^2_{n-1}\ ,
\label{ads}
\ee where $d\Omega^2_{n-1}$ is the metric on a round $S^{n-1}$.
Equation (\ref{stressone}) gives a definition of the action and
stress--tensor on any region (radius $r$ in the coordinates that we
will choose later) bounding the interior of AdS$_{n+1}$. The AdS/CFT
relation equates these quantities to a dual conformal field theory
residing ``on the boundary'' at ($r{\to}\infty$).

As stated before, the metric restricted to the boundary, $h_{ab}$, diverges
due to an infinite conformal factor, which is $r^2/l^2$. We take the
background metric upon which the dual field theory resides as
\be\gamma_{ab}=\lim_{r\to\infty}{l^2\over r^2}h_{ab}\ .
\label{newmetric}
\ee and so the field theory's stress--tensor, ${\widehat T}^{ab}$, is
related to the one in (\ref{stressone}) by the
rescaling\cite{robstress}: \be \sqrt{-\gamma}\,\gamma_{ab}{\widehat
T}^{bc}=\lim_{r\to\infty}\sqrt{-h}\,h_{ab}T^{bc}\ .
\label{newstress}
\ee This amounts to multiplying all expressions for $T^{ab}$ displayed
later by $(r/l)^{n-2}$ before taking the limit $r{\to}\infty$. For the
AdS$_{n+1}$ example (\ref{ads}), the $n$ dimensional boundary upon
which the theory resides is the Einstein universe, with metric
$ds^2=-dt^2+l^2d\Omega^2_{n-1}$. In the case of $n{=}4$, the field
theory stress tensor computed using the above
methods\cite{Balasubramanian} can be written (as for other $n$) in the
standard perfect fluid form\cite{robstress} ($u_a{=}(1,0,0,0)$): \be
{\widehat T}_{ab}={1\over64\pi l G}\left(4u_au_b+\gamma_{ab}\right)=
{N^2\over32\pi^2 l^4}\left(4u_au_b+\gamma_{ab}\right). \ee Here we
used the dictionary\cite{Maldacena} between gravity and field theory
quantities, $G{=}l^3\pi/2N^2$.  The total energy, $E{=}\int\! d^{3}\!x
{\widehat T}_{00}{=}3N^2/16l$, is in fact\cite{Balasubramanian} the
Casimir energy of the ${\cal N}{=}4$ supersymmetric $SU(N)$
Yang--Mills theory on the~$S^3$. (See also ref.\cite{robgary}.)  The
conformal invariance of the theory is evident in the fact that
${\widehat T}_a^a{=}0$. It is worth stressing that there should be no
confusion about the fact that the theory is conformal while in a
box,~$S^3$, which has a scale, $l$. Conformal invariance is preserved
since this scale enters in a conformally invariant way in the metric
itself.

This prescription gives a method for computing the stress tensor of a large
class of field theories, which may be obtained by studying spacetimes which
are asymptotically locally AdS. We will now show that it also provides a
natural method for generating examples of field theories which are scale
but not conformally invariant, by placing a conformal field theory on a
spacetime with just the correct properties we asked for earlier. We shall
present two examples here, and will report more details and examples in an
extended publication\cite{extended}.

Our examples come from the Kerr--AdS spacetimes in five and seven
dimensions (with only one of the two rotation parameters
non--zero\cite{Hawkingtwo}): \bea ds^2&=&-{\Delta_{r} \over
\rho^2}\left(dt-{a \sin^2{\theta} \over \Xi } d\phi\right)^2 +{\rho^2
\over \Delta_{r}}dr^2+{\rho^2\over\Delta_\theta}d\theta^2\nonumber\\ &
&+{\Delta_{\theta}\sin^2{\theta}\over\rho^2}\left(adt-{(r^2+a^2) \over
\Xi} d\phi\right)^2\nonumber\\&&+r^2 \cos^2{\theta} d\Omega_{n-3}^2\
,\nonumber\\\nonumber\\ \Xi&=&1-{a^2/l^2}\ ,\quad
\rho^2=r^2+a^2\cos^2{\theta}\ ,\nonumber\\
\Delta_{r}&=&(r^2+a^2)(1+r^2/l^2)-2MGr^{4-n}\ , \nonumber\\
\Delta_{\theta}&=&1-(a^2/l^2)\cos^2{\theta}\
\label{metricstuff}
\eea where $n{=}4$ or $6$, and: \be d\Omega_1^2=d\psi^2\ ;\,\,\,
d\Omega_3^2=d\psi^2+\sin^2\psi d\eta^2 +\cos^2 \psi d \beta^2\ .  \ee
Using the prescription~(\ref{newmetric}), the metric on which the
field theory (either the ${\cal N}{=}4$ Yang--Mills theory for $n{=}4$
or the (0,2) CFT for $n{=}6$) resides can be seen to be that of a
rotating Einstein universe\cite{cassidy} (see also
refs.\cite{klemm,Berman,harvey,karl}): \bea ds^2&=&
-dt^2+{2a\sin^2{\theta}\over\Xi}dtd\phi+l^2{d\theta^2 \over
\Delta_{\theta}}\nonumber\\ & & \hskip2cm+l^2{\sin^2{\theta} \over
\Xi}d\phi^2+l^2\cos^2{\theta} d \Omega_{n-3}^2\ .
\label{rotatestein}
\eea

The gauge theory stress tensor of the Kerr--AdS$_5$ spacetime was
computed in ref.\cite{adel}, and the full expression can be found
there. In particular the trace is: \be {\widehat T}_{a}^{a}=-{N^2a^2
\over 4 \pi^2l^6}[a^2/l^2 (3\cos^4{\theta}-2\cos^2{\theta})
-\cos{2\theta}]\ .
\label{trace}
\ee

While it is non--zero, a quick computation shows that this is a total
derivative, and it is in fact ${\widehat
T}_{a}^{a}={-}(N^2/\pi^2)E_4$, where the Euler density is: \be
E_4\equiv {1\over64}\left[{\cal R}^{\mu\nu\kappa\lambda}{\cal
R}_{\mu\nu\kappa\lambda} -4{\cal R}^{\mu\nu}{\cal R}_{\mu\nu}+{\cal
R}^2\right] \ . \ee The coefficient is precisely the field theory
value\cite{Henningson}. Since~$\int d^4\!x \sqrt{-\gamma}\, {\widehat T}_{a}^{a}{=}0$,
(which fits that the action, $I$, is computed to be
finite\cite{adel,Hawkingtwo}), this is our first example of having
preserved scale invariance while having broken conformal invariance.
This was considered as a possibility in ref.~\cite{adel}. However, it
was noticed there that the trace was also proportional to $\Box {\cal
R}$. So in fact, it was misidentified there with an anomaly of type
D. A proposal was made there to add an ${\cal R}^2$ counterterm to the
action and define an ``improved''\cite{jackiwone} stress tensor (see
also refs.~\cite{ho}). However, as pointed out in a note--in--proof
added to ref.~\cite{adel}, it was not possible to do this while
preserving the values for the physical conserved quantities (for
example) like angular momentum. The point we stress here is that it is
quite a {\sl special} case that the Euler density happened to be
proportional to $\Box{\cal R}$. In general it cannot be written in
this form. Our anomaly is purely of type A, and as such its
coefficient cannot be changed by adding counterterms. Instead, we
accept the presence of the anomaly and give up conformal invariance;
the rotation parameter $a$ has broken conformal invariance of the
theory, but scale invariance is preserved.

We now turn to the case of Kerr--AdS$_{7}$.  We computed the
non--vanishing components for the stress tensor at large $r$, to
$O(r^{-4})$ using eqns.~(\ref{theterms}) and (\ref{stressone}): \bea
\hskip-1.0cm T_{tt}&=&{l^3 \over 640 \pi G r^4}\left[ (1+a^2/l^2)(-131
a^2/l^2-423a^4/l^4\cos^4{\theta})\right.\nonumber\\ &&+219
(1+a^4/l^4)a^{2}/l^{2}\cos^2{\theta}+25 (1+a^6/l^6)\nonumber\\ &&+235
a^{6}/l^6 \cos^6\theta+400MG/l^4
\left.  +492 a^4/l^4 \cos^2{\theta}\right]+\cdots\ ,\nonumber\\
T_{t\phi}&=&-{l^3 a \sin^2\theta \over 640 \pi Gr^4
\Xi}\left[-231a^{6}/l^6\cos^4\theta+5+a^2/l^2\right.\nonumber\\ &
&+55a^{6}/l^6\cos^6\theta+
189a^{6}/l^6\cos^2{\theta}-25a^6/l^{6}\nonumber\\ &
&+168a^{4}\cos^2{\theta}/l^4-51a^{4}/l^4\cos^4\theta-101a^{4}/l^{4}\nonumber\\
& &\left.-3a^{2}/l^2\cos^2{\theta}+400MG/l^4\right]+\cdots,\nonumber\\
T_{\phi\phi}&=&{al^5 \sin^2\theta \over 640 \pi G r^4
\Xi^2}\left[102a^{4}/l^{4}+51
a^{4}/l^4\cos^4{\theta}\right.\nonumber\\ &
&-55a^{6}\Xi/l^6\cos^6\theta -171a^{4}/l^4 \cos^2\theta+189a^8l^8
\cos^2\theta/\nonumber\\ & &+3a^2/l^2\cos^2{\theta}+4a^{2}/l^{2}-231
a^8/l^8 \cos^4{\theta}\nonumber\\ & &180a^{6}/l^6\cos^4{\theta}+480MG
a^2/l^6\cos^2{\theta}\nonumber\\ & &+80MG/l^4-21a^6\cos^2\theta
/l^6-76a^6/l^6-25a^8/l^8\nonumber\\ & &\left.+400a^2MG/l^6-5\right]+
\cdots\ ,\nonumber\\ T_{\theta\theta}&=&-{l^5 \over 640 \pi G
\Delta_{\theta}r^4}\left[5a^2\Xi/l^2-80MG/l^4\right.\nonumber\\ &
&+3a^2/l^2\cos^2{\theta}(1+a^4/l^4)+66a^4/l^4\cos^2{\theta}\nonumber\\
& &\left.-45a^4/l^4\cos^4{\theta}(1+a^2/l^2)+25a^6/l^6\cos^6\right]
+\cdots\ ,\nonumber\\ T_{\psi\psi}&=&-{l^5\cos^2{\theta} \over 640 \pi
G r^4}\left[5 \Xi(1-a^{4}/l^4)-80MG/l^4\right.\nonumber\\ & &-51
a^{4}/l^4\cos^4{\theta}(1+a^2/l^2)-3a^2/l^2\cos^2\theta(1+a^4/l^4)\nonumber\\
& &\left.+46a^{4}/l^4\cos^2{\theta} \right]+\cdots\ , \nonumber\\
T_{\eta\eta}&=&\sin^2{\psi}T_{\psi\psi},\quad
T_{\beta\beta}=\cos^2{\psi}T_{\psi\psi}.  \eea

A computation shows that the action is again finite: \bea
I&=&{2\pi^3(r_{+}^{2}+a^{2})r_{+}\over G\Xi \left(
3r_{+}^4/l^2+2r_{+}^2(1+a^2/l^2)+a^2\right)}\left[160
(r_{+}^6/l^6\right.\nonumber\\ &
&\left.+r_{+}^4/l^6a^2-M/l^4)+5a^4/l^4+50\Xi+a^6/l^6 \right]\ , \eea where
$r_{+}$ is the location of the horizon, the largest root of $\Delta_r{=}0$.

Taking the trace of the stress tensor yields: \bea {\widehat
T}_a^a&=&-{a^2 N^3 \over 2\pi^3 l^8 }\left[5a^4/l^4 \cos^6\theta -8
\cos^4\theta a^2/l^2(1+a^2/l^2)\right.\nonumber\\ &
&\left.+3\cos^2\theta(1+a^4/l^4+3a^2/l^2)-2(1+a^2/l^2)\right] \eea
where we used the relation\cite{Maldacena} $N^3{=}{3\pi^2 l^5/ 16 G}$
between the gauge theory and the gravity parameters.

Again, we see that this trace is a total derivative. Furthermore,
${\widehat T}_a^a={-}(N^3 / 4508 \pi^3)E_6$. (The Euler density $E_6$
is displayed in {\it e.g.}  ref.\cite{tseytlin}, where the CFT is
discussed at weak coupling). The coefficient matches the results in
refs.\cite{Henningson,kraus,tseytlin}. ({\it n.b.,} a typo in
ref.\cite{Henningson} is corrected in ref.\cite{tseytlin}.) Just as in
the four dimensional case, we see that for this special situation, the
Euler density can be written in terms of type D quantities. In the
notation of ref.\cite{tseytlin}, it is of the form
$\sum_{i=1}^7d_iC_i$, with $d_5$ and $d_7$ zero, since they depend on
the Weyl tensor, and
$\{d_1,d_2,d_3,d_4,d_6\}=\{0,1/9,1/72,-5/12,1/72\}$.  (A useful
parameterisation, but of course, not unique.)  This completes our
second counterexample to the folklore.

In conclusion, we have used the AdS/CFT correspondence to define known
four and six dimensional conformal field theories, at large $N$, on spacetimes with
properties chosen so that conformal invariance, but not scale
invariance, is broken.  The crucial point is that there are ``many
faces''\cite{counter} to anti--de Sitter spacetime, which can be found
by slicing it in different ways by various coordinate choices, and
then picking the boundary for the field theory to live on. The
Kerr--AdS solutions give a particularly interesting slicing, for our
present purposes: The $M{=}0$ limit is just AdS in very non--standard
coordinates. The boundary is related to the standard static Einstein
universe (which has vanishing Euler density) by a complicated change
of variables, and a conformal transformation\cite{Hawkingtwo}.  This
is why the Euler density can be globally written as a total
derivative, while the manifold remains conformally flat.  It would be
interesting to characterise such spacetimes further, since they give a
straightforward means of defining the sort of field theory examples
discussed herein.

\bigskip

{\bf Acknowledgements:} We would like to thank Vijay Balasubramanian,
 Roberto Emparan, Juan Maldacena, Rob Myers, Joe Polchinski, and Al
 Shapere for comments and discussions, and especially Paul Townsend
 for raising the main issue discussed in this paper.  CVJ would like
 to thank the Theory group at Harvard University for hospitality while
 this paper was written.  AMA's work was supported by an NSF Career
 grant, \#PHY--9733173. This paper is report \#'s UK/00--02, and
 DTP/00/43.  This paper was brought to you by the letters A, d, S, C,
 F, and T, and the numbers 4, 5, 6 and 7.

\end{multicols}

\end{document}